\documentclass[sigplan, 10pt]{acmart}
\AtBeginDocument{%
  \providecommand\BibTeX{{%
    \normalfont B\kern-0.5em{\scshape i\kern-0.25em b}\kern-0.8em\TeX}}}

\acmYear{2024}\copyrightyear{2024}
\setcopyright{rightsretained}
\acmConference[EuroMLSys '24]{4th Workshop on Machine Learning and Systems}{April 22, 2024}{Athens, Greece}
\acmBooktitle{4th Workshop on Machine Learning and Systems (EuroMLSys '24), April 22, 2024, Athens, Greece}
\acmPrice{}
\acmDOI{10.1145/3578356.3592577}
\acmISBN{979-8-4007-0084-2/23/05}

\usepackage{caption}
\usepackage{subcaption}

\usepackage{algorithm}
\usepackage{algpseudocode}
\usepackage{amsmath}

\usepackage{listings}
\usepackage{xcolor}
\definecolor{codegreen}{rgb}{0,0.6,0}
\definecolor{codegray}{rgb}{0.5,0.5,0.5}
\definecolor{codepurple}{rgb}{0.58,0,0.82}
\definecolor{backcolour}{rgb}{0.95,0.95,0.92}
\definecolor{codeblue}{rgb}{0,0,1}
\definecolor{codered}{rgb}{1,0,0}

\lstdefinestyle{python}{
    backgroundcolor=\color{backcolour},   
    commentstyle=\color{codegreen},
    keywordstyle=\color{magenta},
    numberstyle=\tiny\color{codegray},
    stringstyle=\color{codepurple},
    basicstyle=\footnotesize,
    breakatwhitespace=false,         
    breaklines=true,                 
    captionpos=b,                    
    keepspaces=true,                 
    numbers=left,                    
    numbersep=5pt,                  
    showspaces=false,                
    showstringspaces=false,
    showtabs=false,                  
    tabsize=2
}

\definecolor{backcolour2}{rgb}{0.92,0.95,0.95}
\definecolor{imadcolor}{rgb}{0.8,0.4,0} 
\definecolor{ldgstscolor}{rgb}{0.4,0,0.8} 

\lstdefinestyle{sass}{
    backgroundcolor=\color{backcolour2},   
    commentstyle=\color{codegreen},
    keywordstyle=\color{codeblue},
    numberstyle=\tiny\color{codegray},
    stringstyle=\color{codered},
    basicstyle=\ttfamily\footnotesize,
    breakatwhitespace=false,         
    breaklines=true,                 
    captionpos=b,                    
    keepspaces=true,                 
    numbers=left,                    
    numbersep=5pt,                  
    showspaces=false,                
    showstringspaces=false,
    showtabs=false,                  
    tabsize=2,
    literate={IMAD}{{\textcolor{imadcolor}{IMAD}}}{4}
    {LDGSTS}{{\textcolor{ldgstscolor}{LDGSTS}}}{6},
}

\usepackage{multicol}

\definecolor{keywordcolor}{rgb}{0.13,0.29,0.53}
\definecolor{commentcolor}{rgb}{0,0.5,0}

\lstdefinelanguage{PTX}{
    morekeywords={mul, lo, s64, shl, b64, add, s32, selp, b32, cp, async, cg, shared, global, commit, group},
    morecomment=[l]{//},
    keywordstyle=\color{keywordcolor}\bfseries,
    commentstyle=\color{commentcolor}\itshape,
}

\lstdefinestyle{myPTXstyle}{
    language=PTX,
    basicstyle=\footnotesize\ttfamily, 
    breaklines=true, 
    columns=fullflexible, 
    numbers=left, 
    stepnumber=1, 
    frame=single, 
    keepspaces=true, 
    captionpos=b, 
    showstringspaces=false, 
}

\begin{document}

\title{SIP: Autotuning GPU Native Schedules via Stochastic Instruction Perturbation}

\author{Guoliang He}
\email{gh512@cam.ac.uk}
\affiliation{
  \institution{University of Cambridge}
  \city{Cambridge}
  \country{United Kingdom}
  }

\author{Eiko Yoneki}
\email{eiko.yoneki@cl.cam.ac.uk}
\affiliation{
  \institution{University of Cambridge}
  \city{Cambridge}
  \country{United Kingdom}
  }

\begin{abstract}

Large language models (LLMs) have become a significant workload since their appearance. However, they are also computationally expensive as they have billions of parameters and are trained with massive amounts of data. Thus, recent works have developed dedicated CUDA kernels for LLM training and inference instead of relying on compiler-generated ones, so that hardware resources are as fully utilized as possible. In this work, we explore the possibility of GPU native instruction optimization to further push the CUDA kernels to extreme performance. Contrary to prior works, we adopt an automatic optimization approach by defining a search space of possible GPU native instruction schedules, and then we apply stochastic search to perform optimization. Experiments show that SIP can further improve CUDA kernel throughput by automatically discovering better GPU native instruction schedules and the optimized schedules are tested by $10$ million test samples.

\end{abstract}




\maketitle

\section{Introduction}
\label{seq. intro}

LLMs are transformer-based deep neural networks (DNNs) consisting of many layers of self-attention \cite{attn} and linear projections. Since their appearance, state-of-the-art performance has been achieved across various domains, such as image generation \cite{sora} and natural language processing \cite{llama}. To date, OpenAI \cite{sam_twitter, chatgpt} announces about 100 billion words are generated every day. As such, LLMs have become a significant workload in the deep learning community and have gathered much attention. 

However, training and serving LLMs are computationally expensive because they typically consist of multiple layers of transformer backbone, which is billions of parameters. As a result, researchers have explored various ways to accelerate LLM computation. For example, fused attention (flash-attention) \cite{flash_attn} is developed such that the attention computation achieves better utilization of the shared memory of NVIDIA GPUs, and therefore it is faster than compiler-generated kernels. Paged attention is an efficient algorithm to manage KV caching \cite{vllm} and avoid GPU memory fragmentation. We observe that those works are typically implemented by hand-written hardware-efficient codes, i.e. CUDA kernels for NVIDIA GPUs, for the flexibility and efficiency of hardware-vendor-provided programming models. 

In this work, we investigate the possibility of further improving the hand-written kernels by exploring optimization at a lower level, i.e. hardware native assembly. Specifically, we focus on NVIDIA CUDA kernels. CUDA kernels are usually programmed in C/C++ syntax and are compiled to \textit{ptx} \cite{cuda_ptx}, an intermediate virtual assembly. Then, \textit{ptx} is lowered to \textit{sass}, which is native to the target GPU architecture. Finally, the \textit{sass} codes are compiled into binary codes (\textit{cubin}) that can be executed directly on the GPU. We observe that the native \textit{sass} schedule is not accessible when programming in C/C++ syntax due to the multi-stage compilation, and thus certain optimization opportunities are locked.

We propose SIP, the first automatic optimizer for optimizing \textit{sass} schedules. SIP takes as input a \textit{cubin} (compiled from a kernel) and then performs a stochastic search to optimize the schedules, and finally outputs an optimized \textit{cubin}. An autotuning approach is taken, as manual scheduling is tedious and error-prone, as well as cannot cope with the nearly infinite configuration space. Therefore, SIP is capable of automatically discovering better schedules for a given \textit{cubin}.

In summary, this paper makes the following contributions:

\begin{itemize}
    \item We introduce SIP, an automatic optimizer for optimizing NVIDIA GPU native schedules.
    \item Our evaluation covers two representative workloads for LLMs and shows that SIP can further improve the throughput of existing GPU kernels.
\end{itemize}

\section{Background and motivation}

\begin{figure*}[htb]
    \centering
    \includegraphics[width=0.5\linewidth]{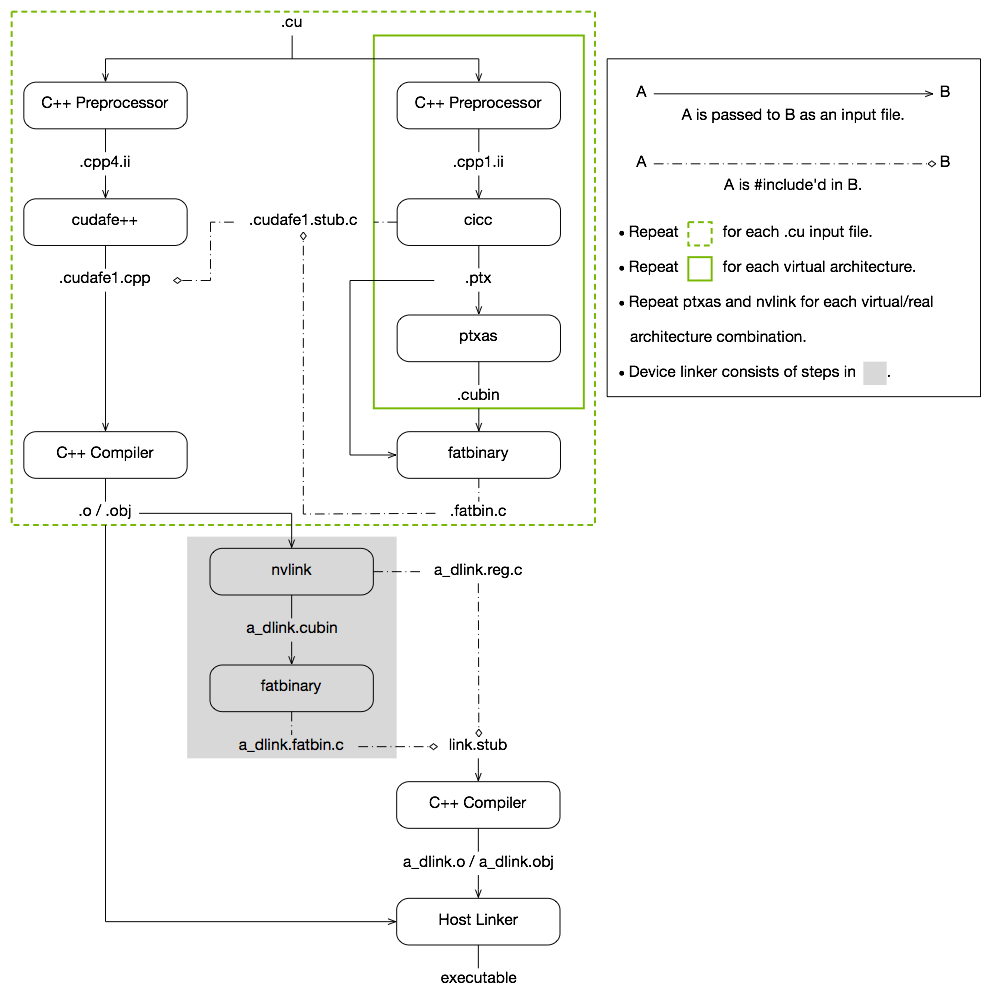}
    \caption{CUDA compilation as documented by NVIDIA \cite{cuda_compiler}. The solid green circle indicates the compilation of device (CUDA) codes.}
    \label{fig: Cuda compilation}
\end{figure*}

\subsection{Programming GPUs}

GPUs are multi-threaded hardware accelerators that can perform highly parallel computation and therefore tensor operations such as matrix multiplication can be executed efficiently. However, it is the programmer's responsibility to map the tensor operation to the hardware. As a result, programmers must follow the programming model provided by CUDA \cite{cuda}. Conceptually, a CUDA kernel consists of a grid of thread blocks running concurrently, and inside each thread block are multiple threads. Each thread block is mapped to a GPU steaming multiprocessor and is executed individually in parallel.

\subsection{Compiling CUDA codes}

Although CUDA kernels have a syntax that looks like a C++ function, they must be compiled by the proprietary compiler (\textit{NVCC}) provided by NVIDIA. The compilation process has several stages: first, the CUDA code is compiled to \textit{ptx}, which is an intermediate language that is GPU-architecture independent \cite{cuda_ptx}. Note that one can also directly program \textit{ptx} and it is the lowest-level programming interface supported by NVIDIA officially. 

Then the \textit{ptx} codes are lowered to \textit{sass}, which is native to the target GPU. That is, for different GPU architectures, the generated \textit{sass} codes are different even though the original \textit{ptx} codes are identical. While the corresponding \textit{sass} codes of a CUDA kernel are obtainable by utilizing the CUDA binary utilities \cite{cuda_sass}, the instruction set is only vaguely documented. As a result, it is unclear the lowering and optimization process at this stage. 

Finally, the \textit{sass} codes are compiled into binary code (\textit{cubin}) that can be executed directly on the GPU. The overall compilation process is shown in Figure \ref{fig: Cuda compilation}.


\subsection{Programming GPUs-native instructions} 
\label{sec. native instructions}

Prior works have shown that programming at the GPUs-native level can unlock optimization opportunities \cite{opt_conv_gpu,opt_gemm_gpu}. This is because direct control of registers and instruction schedules is not accessible when programming at a higher level. Moreover, since the \textit{Kepler} GPU architecture \cite{cuda_kepler}, it seems that the GPU hardware obeys the compiler-generated \textit{sass} instructions at runtime. This means the static instruction schedules of GPUs are not executed out-of-order by the hardware. As a result, a better static instruction schedule directly leads to better performance.

\subsubsection*{Latency hiding.} Prior works \cite{opt_conv_gpu,opt_gemm_gpu} show a methodology for hiding latency by manually reordering native instructions, which overlaps the memory I/O and computation instructions as much as possible. Specifically, the instruction execution is stalled when its dependent data are not ready. This is because memory instructions such as reading and loading data from global memory do not have a fixed latency due to the deep hierarchy of GPU memory systems. As a result, the data dependencies must be resolved by the compiler through the control code attached to the instruction \cite{dissect_turing, cuda_cuasm}. For example, a \textit{sass} instruction may look like:

\begin{verbatim}
      [B------:R-:W2:-:S02] LDG.E R0, [R2.64] ;
\end{verbatim}

The control code is enclosed by square brackets and is separated into multiple fields by colons \cite{cuda_cuasm}. The first field is the wait barrier mask, if any of the bits are set, the instruction is stalled until the bit is clear. The second and third fields are read and write barrier masks. In this case, this instruction sets the write barrier to 2, which means a later instruction whose data depending on $R0$ is stalled until $R0$ is ready.

In this case, if the next instruction depends on $R0$, the instruction may be stalled and incur latency, as the compiler will set the wait barrier to $2$. Therefore, it is more beneficial to separate the two instructions further away to hide the latency and overlap the memory I/O and computation instructions as much as possible.

\subsubsection*{Challenges.} However, programming at the native instruction level is extremely challenging for several reasons. Firstly, programmers must resolve the data dependencies between \textit{sass} instructions manually. Secondly, the native instructions (\textit{sass}) are not well documented and vary across GPU architectures. Lastly, there is not any official assembler that can be used to assemble the \textit{sass} into binary codes. The lack of official assemblers in turn makes it hard to understand \textit{sass} instructions and perform microbenchmarks. Fortunately, the open-source community has implemented unofficial assemblers to assemble the \textit{sass} into binary codes. For example, there are \textit{MaxAs} \cite{cuda_maxas}, \textit{TuringAs} \cite{opt_conv_gpu}, and \textit{Cuasm} \cite{cuda_cuasm}. In this work, we leverage \textit{Cuasm} to assemble the \textit{sass} into binary codes.

\subsection{Necessity for automatic optimization}

\begin{table}[htb]
  \centering
  \caption{Example of circles per instruction (CPI) for Pascal, Volta, and Turing GPUs \cite{dissect_turing}.}
  \begin{tabular}{|c|c|c|}
    \hline
    GPU & Inst. & CPI \\
    \hline
    Pascal & FFMA & 6 \\
     & POPC & $\sim14$ \\
     & IMAD & $\sim86$ \\
    \hline
    Volta & FFMA & 4 \\
     & POPC & $\sim10$ \\
     & IMAD & 5 \\
    \hline
    Turing & FFMA & 4 \\
     & POPC & $\sim15$ \\
     & IMAD & 5 \\
    \hline
  \end{tabular}
  \label{tab. inst CPI}
\end{table}

In this work, we make the first attempt to automatically optimize the \textit{sass} schedule. While prior works have shown methodologies for optimizing the \textit{sass} schedule, it requires in-depth knowledge and microbenchmarks to obtain the latency of the \textit{sass} instructions, which is tedious, and not to mention resolving the data dependencies between \textit{sass} schedules.

Moreover, the instruction latencies vary across GPU architectures. As shown by Table \ref*{tab. inst CPI}, the CPI of some of the \textit{sass} instructions can be significantly different (up to $17.2$x) across GPU architectures. Therefore, the expert schedule may be ineffective for newer-generation GPUs.


\section{SIP}

In this section, we introduce SIP, an automatic optimizer for GPU native instruction schedules. One of the possible approaches to achieve automatic optimization is through search-based optimization, that is, we design a search space and then use a search algorithm to explore the search space. This effectively forms a control loop and lets us trade off the search time and optimized performance. In the following section, we define the necessary components to form a control loop: the search space, the mutation policy, the feedback signal, and the search algorithm.


\subsection{Search space} 

The complete search space is the full permutation of the \textit{sass} instructions. Given $n$ instructions, we could generate $n!$ possible schedules by enumerating the permutation. However, a typical GPU kernel often consists of thousands of instructions, which means the potential schedules are more than the number of atoms in the known universe. Therefore, it is impossible to find the optimal one within a reasonable time.

\subsubsection*{Pruning.} We therefore prune the search space by limiting our choices. Considering that latency hiding optimization is mostly achieved by interleaving memory I/O instructions and compute instructions, we only consider the memory I/O instructions in the search space. Furthermore, accessing the global memory is the most expensive memory I/O operation, and therefore we only consider global memory read and write. In our experiment, this successfully reduces the number of considered instructions to a few hundred. Pruning represents our trade-off between search time and the optimization space.

\subsection{Mutation policy} 
\label{sec:mutation_policy}

Having defined the search space, we design a mutation policy that allows us to explore it. Considering we would like to overlap the memory I/O and computation as much as possible, we allow each memory I/O instruction to move up or down by one in the schedule at each iteration. Intuitively, this corresponds to experts reordering the instruction schedule.

Specifically, if there exist $k$ memory I/O instructions, the mutation policy may choose one of them to move up or down by one. The exact instruction to move and direction is randomly chosen. The action vector is two discrete numbers, where the first one represents which memory I/O instruction to move and the second one represents the direction.

\subsection{Feedback signal}
\label{sec:feedback_signal}

Obtaining the feedback signal is the most important component as it directly guides the search algorithm to explore the search space. In this work, we mostly care about the runtime of the optimized CUDA kernels, and therefore we must measure the runtime after the mutation policy is applied. Measuring the runtime can be achieved by either cost modeling or actual execution. 

\subsubsection*{Cost modeling.} While there exists a GPU simulator from the open-source community \cite{gpgpu_sim} that could provide the feedback signal, we are concerned that it is not actively maintained. Therefore, it may not simulate the latest GPU hardware. Moreover, we do not know the time needed to perform a simulation. Therefore, we use the actual execution approach in this work.

\subsubsection*{Execution on GPUs} As mentioned in \S\ref{sec. native instructions}, we use an open-source assembler \textit{Cuasm} to assemble the mutated \textit{sass} into binary codes, and then we use CUDA events to measure the execution time by first warm up the GPU and then measure the elapsed time. We use the following formula to obtain the feedback signal:

\begin{equation}
  R = \frac{T_{i-1} - T_i}{T_0}
\end{equation}

Where $T_0$ is the initial runtime, $T_i$ is the runtime after the mutation policy is applied, and $T_{i-1}$ is the runtime before the mutation policy is applied. Intuitively, this gives positive feedback if the mutation policy decreases the runtime, and negative feedback if the mutation policy increases the runtime.


\subsection{Search algorithm}
\label{sec:search_algorithm}

The search algorithm forms the control loop by connecting the other components and it is detailed in algorithm \ref{alg:search_algorithm}. 

\begin{algorithm}
  \caption{Simulated annealing for stochastic instruction Perturbation (SIP).}
  \begin{algorithmic}[1]
  \State Initialize $T_{\text{max}}$, $T_{\text{min}}$, $x$ 
  \State Initialize $x_{\text{best}} \gets x$ 
  \State $T \gets T_{\text{max}}$ 
  
  \While{$T > T_{\text{min}}$}
        \State Generate a new schedule $x'$ by perturbing $x$
        \State $\Delta E = \text{Energy}(x') - \text{Energy}(x)$
        \If{$\Delta E < 0$}
            \State Accept $x'$ as the current schedule: $x \gets x'$
            \If{$\text{Energy}(x) < \text{Energy}(x_{\text{best}})$}
                \State Update the best schedule: $x_{\text{best}} \gets x$
            \EndIf
        \Else 
            \If{$r < \exp(-\Delta E / T)$}
                \State Accept $x'$ as the current schedule: $x \gets x'$
            \EndIf
        \EndIf
      \State Cool down the system: $T \gets T \times L^{-1}$
  \EndWhile
  \State \textbf{return} $x_{\text{best}}$ 
  \end{algorithmic}
  \label{alg:search_algorithm}
  \end{algorithm}

The measurement of energy is based on the feedback signal \S\ref{sec:feedback_signal}, and the perturbation is based on the mutation policy \S\ref{sec:mutation_policy}.

\section{Implementation}
\label{sec. impl}

\subsection{Integration to Triton}

Technically, SIP can be applied to arbitrary GPU kernels. It simply takes as input a compiled kernel (\textit{cubin}), and then performs the search-based optimization. The output will be an optimized \textit{cubin}.

For evaluation, we choose to integrate with OpenAI Triton \cite{triton}, which is a compiler for writing GPU kernels. Triton allows users to write kernel codes in \textit{Python} syntax and then jit-in-time compile to either NVIDIA GPUs or AMD GPUs. As a result, it greatly simplifies the process of testing research ideas and implementing high-performance GPU kernels for the machine learning community. Moreover, Triton is also the default backend of Pytorch \cite{pt2}, which is one of the most popular deep-learning frameworks. By integrating with Triton, we hope our work can benefit the deep-learning community directly. 

\begin{lstlisting}[language=Python, caption=Example Triton kernel codes, style=python]
  @triton.jit
  def vector_add(x_ptr, out_ptr):
    ...
\end{lstlisting}

SIP reuses Triton's compilation pipeline but intercepts the compiled \textit{cubin}. It then applies \textit{cuasm} \cite{cuda_cuasm} to disassemble the \textit{cubin} into \textit{sass} and extracts the native instruction of the kernel while keeping the other meta-information intact. This is important as the meta-information such as the symbol tables and the ELF format must be preserved. Finally, it applies stochastic search and substitutes the kernel section with the optimized \textit{cubin}. To invoke SIP's optimization, users simply need to change one line in the Triton code.

\begin{lstlisting}[language=Python, caption=SIP example, style=python]
  @sip.jit(ret_ptr=1)
  def vector_add(x_ptr, out_ptr):
    ...
\end{lstlisting}

Where the \textit{ret\_ptr} is the index to the output buffer and is used in the testing (\S\ref{sec. testing}). Optionally, users may add more arguments to specify the search budget, hyperparameters, etc. 

As SIP performs a stochastic search, the optimized performance and search budget may be indeterministic. The optimized performance also depends on the search budget as a larger search budget means more search space is explored. Therefore, SIP is expected to perform offline searches and store results from multiple rounds of searches. Then it applies a greedy algorithm to rank all found \textit{cubin} and picks the best one if it passes all tests. Finally, at deployment, i.e. training LLMs, the best \textit{cubin} is retrieved and loaded into Triton directly without incurring any runtime overhead. In SIP, this can be achieved by passing in an argument to specify the path to the pre-selected \textit{cubin}.

\subsection{Automatic probabilistic testing}
\label{sec. testing}

Mutation of the \textit{sass} causes transformation correctness concerns. After all, the results will be meaningless if the kernel computation is incorrect. There are two forms of commonly adopted approaches to evaluate the transformation correctness for assembly codes: probabilistic testing and validation. 

Probabilistic testing uses reference inputs and outputs to test the correctness of the mutated program. It is relatively cheap and therefore is employed at each step of the search procedure. Given the compiler hint \textit{ret\_ptr}, SIP intercepts the input arguments and randomly generates reference input for tensor data. It then executes the kernel to obtain the reference output. If the reference output does not match the mutated output, a $0$ feedback signal is generated.

Validation involves the use of a theorem solver to mathematically prove the mutated assembly sequences produce the same machine states. For example, the peehole superoptimizer \cite{peehole_opt} uses a symbolic executor and SAT solver to perform validation. However, validation is impossible for GPU native assembly codes because the formal semantics of the \textit{sass} is closed-source. On the other hand, GPU kernels are executed as a C++ function with input and output pointers passed in. Therefore, we can only perform extensive probabilistic testing on the optimized kernels as an approximation of end-to-end equivalency.




%

\section{Evaluation}

In this section, we aim to evaluate SIP to answer the following questions:

\begin{itemize}
  \item Can SIP improve the performance of hand-written CUDA kernels by optimizing at the \textit{sass} level?
  \item What is the relationship between test samples and the number of false positive kernels?
  \item Is it possible to get the same optimization results at a higher-level programming interface?
\end{itemize}


\subsubsection*{Experiment setup} We evaluate SIP with an NVIDIA A100 GPU (Ampere). We use the default NVIDIA compiler \textit{NVCC} $12.2$ and Triton v$2.1.0$. Triton compiles the kernel codes into \textit{ptx} and then \textit{ptxas} compiles to \textit{cubin}. The default \textit{ptx} optimization level is $O3$. 


We choose to evaluate SIP on two representative kernels for LLMs, fused attention (flash-attention) and fused GEMM LeakyReLU. The Triton implementation of the two workloads is available from its official GitHub repository \cite{triton_github}. For all experiments, we use half-precision (\textit{float16}) tensors.

\subsection{Kernel statistics}

\begin{table}[ht!]
  \caption{Fused attention on A100. The input data have the format of $[1, 4, 16384, 64]$, representing \textit{batch size, number of heads, sequence length, head dimension}. }
  \begin{tabular}{p{3cm}p{1.2cm}p{1.2cm}p{1.2cm}}
  \toprule
  \textbf{Metric} & \textbf{Unit} & \textbf{SIP} & \textbf{Triton} \\
  \midrule
  DRAM Frequency & cycle/ns & 1.51 & 1.49 \\
  SM Frequency & cycle/ns & 1.06 & 1.05 \\
  Memory Throughput & \% & \textbf{33.19} & 31.38 \\
  DRAM Throughput & \% & 1.12 & 1.07 \\
  Duration & ms & \textbf{1.29} & 1.37 \\
  L1/TEX Cache  & \% & 44.46 & 44.69 \\
   Throughput &  &  &  \\
  L2 Cache & \% & 16.29 & 15.41 \\
  Throughput &  &  &  \\
  Compute (SM)  & \% & \textbf{45.87} & 43.39 \\
  Throughput &  &  &  \\
  \bottomrule
  \end{tabular}
  \label{tab. speed of light}
\end{table}

We use Nsight Compute \cite{ncomp}, NIVDIA's primary profiler, to study the hardware metrics of the optimized kernels from SIP and Triton respectively. Nsight Compute can be used to extract fine-grained statistics of the optimized kernels with access to NVIDIA's GPU performance counter \cite{nv_perfcounter}. Speed of light (SoL) is the first section reported by Nsight Compute, indicating the achieved performance versus the theoretical maximum. Table \ref{tab. speed of light} shows the results. 


We can see the overall execution duration of the fused attention kernel of SIP ($1.29$ms) is $6.2\%$ lower than Triton ($1.37$ms). This is attributed to both the Compute Throughput being higher ($5.7\%$) and the memory throughput being higher ($5.8\%$). The memory throughput improvement appears to be mostly coming from the L2 cache ($16.29$ versus $15.41$), and the DRAM Throughput comparison also confirms the same, as DRAM is associated with the L2 cache. The utilization of the L1 cache seems to be the same. This is consistent with prior work \cite{opt_conv_gpu}, because their results show a larger speedup can be achieved by reordering global memory instruction compared to local memory instruction.

\begin{table}[ht!]
  \caption{Fused GEMM LeakyReLU on A100. The input data have the format of $[512, 512, 2048]$ (M, N, K).}
  \begin{tabular}{p{3cm}p{1.2cm}p{1.2cm}p{1.2cm}}
  \toprule
  \textbf{Metric} & \textbf{Unit} & \textbf{SIP} & \textbf{Triton} \\
  \midrule
  DRAM Frequency & cycle/ns & 1.50 & 1.50 \\
  SM Frequency & cycle/ns & 1.01 & 1.02 \\
  Memory Throughput & \% & \textbf{45.64} & 40.52 \\
  DRAM Throughput & \% & 9.14 & 8.14 \\
  Duration & $\mu$s & \textbf{23.97} & 26.91 \\
  L1/TEX Cache  & \% & 38.16 & 37.94 \\
   Throughput &  &  &  \\
  L2 Cache & \% & 45.64 & 40.52 \\
  Throughput &  &  &  \\
  Compute (SM)  & \% & \textbf{19.98} & 17.73 \\
  Throughput &  &  &  \\
  \bottomrule
  \end{tabular}
  \label{tab. GEMM speed of light}
\end{table}

A similar pattern for the GEMM LeakyReLU kernel can be observed from Table \ref{tab. GEMM speed of light}, where the SIP-optimized schedule achieves a $12.27\%$ lower latency, and both the compute and memory throughput are higher. 

\subsection{Transformation correctness}

In this section, we evaluate the number of test samples versus the number of incorrect kernels being detected. Full coverage of all possible input data (permutating all bits) is computationally intractable. 

\begin{figure}[htb]
  \centering
  \includegraphics[width=\linewidth]{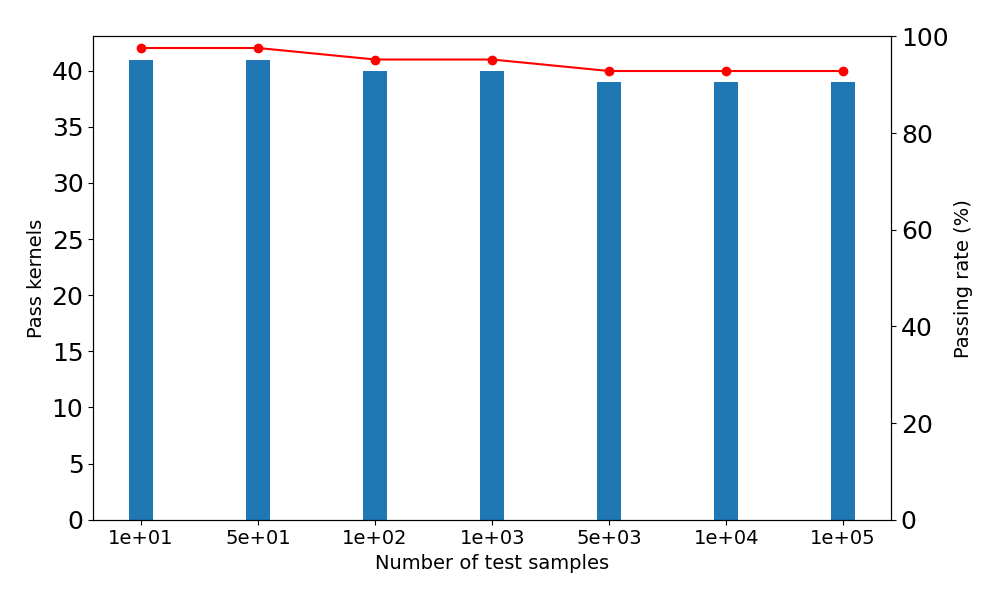}
  \caption{The left y-axis and the bar plot show the number of optimized kernels that pass all tests. The right y-axis and the line plot show the percentage of the successfully tested kernels.}
  \label{fig: test}
\end{figure}

  \vspace{-2mm}

Figure \ref{fig: test} shows the number of kernels that pass all test cases vs. the number of test samples. Initially, $2$ false positive kernels pass the test samples. However, from around $5000$ test samples, the number of optimized kernels that pass all tests becomes stable. 

For the kernels that are reported by Table \ref{tab. speed of light} and \ref{tab. GEMM speed of light}, we ran $10$ million test samples and found all test cases were passed. As the training of llama 2 \cite{llama} uses $2$ trillion tokens and $4$ million batch sizes, which corresponds to $500,000$ forward execution, we consider the test samples to be enough for training purposes. Testing of $10$ million samples costs about $10$ hours but this process can be fully parallelized if multiple GPUs are available. 

\subsection{Necessity for native instruction programming}

In this section, we investigate the necessity of performing optimization at the native instruction level. Specifically, we compare the \textit{ptx} code (the official programming interface at the lowest level) and \textit{sass} instructions. Note that \textit{sass} varies across GPU architectures, so the same \textit{ptx} will produce different \textit{sass} codes for different GPU architectures. 

Considering \textit{ptx} code snippet in Listing \ref{lst:ptx_example}, where a sequence of operations is performed to calculate the address and to load data from global memory to shared memory.

The corresponding \textit{sass} is listed in Listing \ref{lst. triton_sass_example}. Note that the consecutive \textbf{\textit{cp.async}} (in \textit{ptx}) is translated to \textbf{\textit{LDGSTS}} (native \textit{sass} for Ampere GPUs) and interleaved with address calculation automatically by the compiler (\textit{ptxas}'s $O3$ optimization). However, the SIP optimized \textit{sass} reorders certain sequences of the instruction, as shown in Listing \ref{lst. sip_sass_example}, which is not accessible in the \textit{ptx} programming interface. This demonstrates the necessity of \textit{sass}-level optimization, which has direct control over the native instruction schedule.

Unfortunately, due to the nature of the stochastic optimization algorithm, this kind of reordering takes place across the assembly file and it is difficult to understand which reordering sequence is the most significant. We leave it for future work.


\section{Limitation and future work}

The limitation of SIP is that its search algorithm may be ineffective. Reordering the instruction schedule is essentially a high-dimensional discrete optimization problem, and simulated annealing is unable to explore the search space efficiently given the feedback signal. For future work, better search algorithms such as Reinforcement Learning may be applied to enhance SIP, but training RL agents can also be expensive. We also aim to study the optimized \textit{sass} schedule and further understand which reordering may be the most significant to improve performance. We plan to evaluate more GPU architectures too, as different GPU architectures have different performance characteristics.



\section{Conclusion}

We introduce SIP, the first automatic optimizer for GPU native instruction schedules. SIP performs optimization at the GPU native instruction level so thus it is capable of further automatically improving the performance of existing hand-written CUDA kernels. We show that the common compute kernels in LLMs can be improved by around $10\%$ compared to the state-of-the-art. 



\bibliographystyle{ACM-Reference-Format}
\bibliography{sample-base}

\appendix

\section{ptx and sass examples}

%
\begin{lstlisting}[style=myPTXstyle, caption={PTX Code Example}, label=lst:ptx_example]
  add.s64      %rd86, %rd19, %rd142;
  shl.b64      %rd143, %rd139, 1;
  add.s64      %rd87, %rd19, %rd143;
  shl.b64      %rd144, %rd140, 1;
  add.s64      %rd88, %rd19, %rd144;
  shl.b64      %rd145, %rd141, 1;
  add.s64      %rd89, %rd19, %rd145;
  add.s32      %r119, %r204, 16384;
  add.s32      %r121, %r204, 18432;
  add.s32      %r123, %r204, 20480;
  add.s32      %r125, %r204, 22528;
  selp.b32     %r120, 16, 0, %p10;
  cp.async.cg.shared.global [ %r119 + 0 ], [ %rd86 + 0 ], 0x10, %r120;
  cp.async.cg.shared.global [ %r121 + 0 ], [ %rd87 + 0 ], 0x10, %r120;
  cp.async.cg.shared.global [ %r123 + 0 ], [ %rd88 + 0 ], 0x10, %r120;
  cp.async.cg.shared.global [ %r125 + 0 ], [ %rd89 + 0 ], 0x10, %r120;
  cp.async.commit_group ;
\end{lstlisting}

\begin{figure*}[!ht]

\begin{multicols}{2}
  [
  Comparison of SASS Assembly snippets. Note the differences in opcode and register usage.
  ]
  
  \begin{lstlisting}[caption=Triton, style=sass, label=lst. triton_sass_example]
  LDGSTS.E.BYPASS.128 [R219+0x4000], desc[UR16][R10.64], P0 ;
  IMAD.WIDE R18, R9, 0x80, R10 ;
  LDGSTS.E.BYPASS.128 [R219+0x4800], desc[UR16][R44.64], P0 ;
  IMAD.WIDE.U32 R16, R222, 0x2, R64 ;
  LDGSTS.E.BYPASS.128 [R219+0x5000], desc[UR16][R46.64], P0 ;
  IMAD.WIDE.U32 R10, R222, 0x2, R60 ;
  LDGSTS.E.BYPASS.128 [R219+0x5800], desc[UR16][R48.64], P0 ;
  MOV R33, c[0x0][0x1b0] ;
  LDGDEPBAR ;
  \end{lstlisting}
  
  \columnbreak
  
  \begin{lstlisting}[caption=SIP, style=sass, label=lst. sip_sass_example]
  IMAD.WIDE R18, R9, 0x80, R10 ;
  LDGSTS.E.BYPASS.128 [R219+0x4000], desc[UR16][R10.64], P0 ;
  LDGSTS.E.BYPASS.128 [R219+0x4800], desc[UR16][R44.64], P0 ;
  IMAD.WIDE.U32 R16, R222, 0x2, R64 ;
  LDGSTS.E.BYPASS.128 [R219+0x5000], desc[UR16][R46.64], P0 ;
  LDGSTS.E.BYPASS.128 [R219+0x5800], desc[UR16][R48.64], P0 ;
  IMAD.WIDE.U32 R10, R222, 0x2, R60 ;
  MOV R33, c[0x0][0x1b0] ;
  LDGDEPBAR ;
  \end{lstlisting}
  
  \end{multicols}
\end{figure*}

%
%
%
%
%
%
%
%

\end{document}